\documentclass[twoside]{dis08}
\usepackage[latin1]{inputenc}
\usepackage[dvips]{graphicx,epsfig,color}
\usepackage{wrapfig,rotating}
\usepackage{amssymb,amsmath,array}

\begin{document}
\title{Photon Physics at LHC}

\author{M. Vander Donckt
%
\thanks{now at CERN, 1211 Gen\`eve 23, Switzerland}
%
\vspace{.3cm}\\
%
CP3 - Universit\'e Catholique de Louvain, 1348 Louvain-La-Neuve, Belgium
}

\maketitle

\begin{abstract}
Experimental prospects for studying high-energy photon-photon and photon-proton interactions at the \textsc{lhc} are discussed. 
Assuming a typical \textsc{lhc} multipurpose detector, various signals and their irreducible backgrounds are presented after applying acceptance cuts. Selection strategies based on photon interaction tagging techniques are presented.
Prospects are discussed for the Higgs boson search, detection of \textsc{susy} particles and of anomalous quartic gauge couplings, as well as for the top quark physics.
\end{abstract}
\section{Introduction}
\begin{wrapfigure}{r}{0.42\columnwidth}
\vspace*{-0.6cm}
\epsfig{file=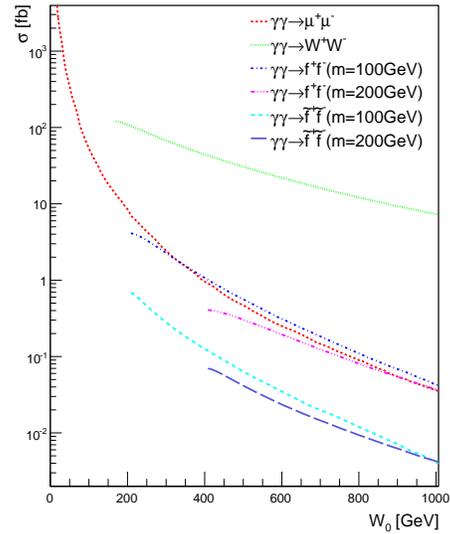,width=\linewidth}
\caption{\footnotesize Cross sections for various $\gamma \gamma$ processes as a function
 of the minimal photon-photon c.m.s. energy $W_0$. No cut is applied}\label{Fig:gammagammaxsec}
\end{wrapfigure}
A significant fraction of $pp$ collisions at the \textsc{lhc} will involve (quasi-real) photon interactions, at center-of-mass (c.m.s) 
energies well beyond the electroweak energy scale. Hence, the \textsc{lhc} can to some extend be considered 
as a high-energy photon-photon or photon-proton collider.

The equivalent photon approximation (\textsc{epa}) can be successfully used to describe the majority of processes involving 
photon exchange, provided that the amplitude of a given process can be factorised into the photon exchange and 
interaction parts~\cite{EPA}.  The photon-photon and photon-proton cross sections, $\sigma_{\gamma\gamma}$ and $\sigma_{\gamma p}$, 
must be convoluted with the photon spectra \sloppy $dN(E_\gamma,Q^2, E)$ to obtain the $pp$ cross sections.
This paper considers the low pile-up conditions available at start-up and focuses on the irreducible backgrounds to the presented analyses. Deeper studies involving inclusive background will be the object of later communications.

\section{Photon-Photon Interactions}
Many $\gamma \gamma$ cross sections at the \textsc{lhc} are sufficiently large
to yield interesting measurements (see figure \ref{Fig:gammagammaxsec}). In particular, the expected high statistics for exclusive
$W$~pair production should allow precise measurement of the $\gamma\gamma WW$ quartic couplings. The
existence of new massive charged particles can also be probed using the two-photon pair production. The 
particular case of super-symmetric pairs is considered here. 
Moreover, the two-photon exclusive production of muon pairs will be an excellent luminosity monitoring tool~\cite{lumi}.

Two-photon exclusive production of muon pairs at the \textsc{lhc} has a well known cross section, and requires very small hadronic corrections. Small theoretical uncertainties and a large cross section ($\sigma = 72.5$~pb at \textsc{lhc} energies) makes it a perfect candidate for the \textsc{lhc} absolute luminosity measurement \cite{lumi}. The selection procedure is very simple: two opposite charge muons within the central detector acceptance ($|\eta|<2.5$), with transverse momenta above some low thresholds ($p_T^{\mu}>3$ or $10$~GeV). Requiring one proton tagged would yield  $150$ muon pairs detected in a $12$ h run at the average luminosity of $5\times10^{32}~\textrm{cm}^{-2}\textrm{s}^{-1}$, allowing run by run calibration of the very forward detectors (\textsc{vfd}).

Two-photon production of $W$ boson pairs provides a unique opportunity to investigate anomalous gauge boson couplings, in particular the quartic gauge couplings (\textsc{qgc}) $\gamma\gamma WW$. For 1~$\textrm{fb}^{-1}$, the obtained limits of $0.49$, $1.84$, $0.54$ and $2.02 \times 10^{-6}$ for $\mathrm{|a^Z_0/\Lambda^2|}$, $\mathrm{|a^Z_C/\Lambda^2|}$, $\mathrm{|a^W_0/\Lambda^2|}$, $\mathrm{|a^W_C/\Lambda^2|}$, respectively, are about 10,000 times better than the best limits established at \textsc{lep2}
\cite{OPAL.limits} clearly showing large and unique potential of such studies at the \textsc{lhc}. The unique signature with large lepton acoplanarity (and large missing transverse momentum) drastically reduces the backgrounds.

\begin{wrapfigure}{r}{0.42\columnwidth}
\begin{center}
\vspace*{-1.0cm}
\epsfig{file=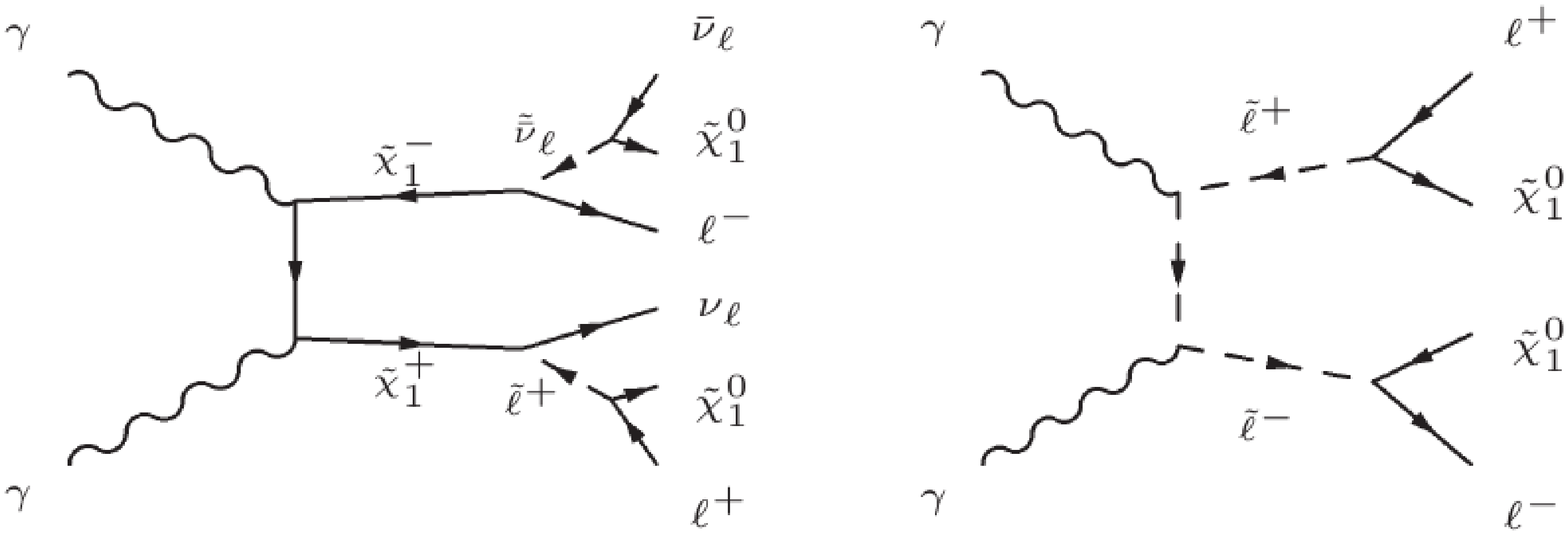,width=\linewidth}
\caption{\footnotesize Most relevant Feynman diagrams for \textsc{susy} pair production with leptons in
 the final state. Left: chargino decay in a charged/neutral scalar and a
 neutral/charged fermion; Right: slepton decay.}
\label{fig.diag.susy}
\end{center}
\end{wrapfigure}

The interest in the two-photon exclusive production of pairs of new charged particles is three-fold: it provides a new, complementary and very simple production mechanism;  constraints on the masses of new particles could be obtained using double \textsc{vfd} tagged events; finally, final states involving super-symmetric particles are 
usually produced without cascade decays. The fully leptonic final state consists of two acoplanar charged leptons and large missing energy has low background.
The corresponding Feynman diagrams are shown at figure \ref{fig.diag.susy}.

The only irreducible background for this type of processes is the exclusive $W$ pair 
production. Indeed, direct lepton pairs $pp(\gamma \gamma \to \ell^+ \ell^-)pp$ can be suppressed using large acoplanarity cuts. Discovery of super-symmetry via $\gamma\gamma$ interaction is \emph{a priori} difficult because of production cross sections smaller than 1~$\textrm{fb}$ before acceptance cuts. However, the measured energy of the two scattered protons in \textsc{vfd}s could allow for the distinction between various contributions to the photon-photon invariant mass $W_{\gamma \gamma}$ \cite{Nicolas}.

\section{Photon-Proton Interactions}
\begin{wrapfigure}{r}{0.45\columnwidth}
\vspace{-0.5cm}
\centerline{\includegraphics[width=0.45\columnwidth]{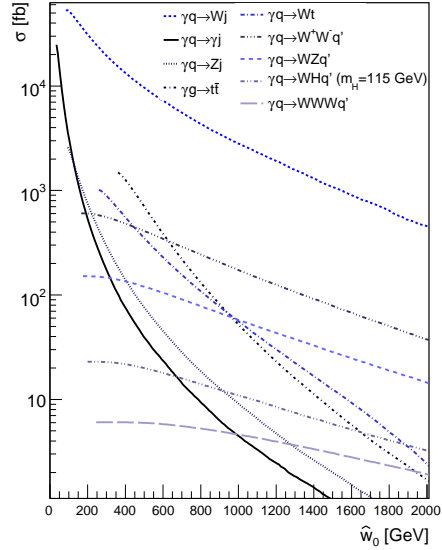}}
\caption{\footnotesize Cross sections for $pp (\gamma q/g \rightarrow X) p Y$ 
processes as a function of the minimal photon-parton c.m.s. energy $\hat{W}_0$.}\label{Fig:gammapartonxsec}
\end{wrapfigure}

The high luminosity and the high c.m.s. energy of photo-production processes at \textsc{lhc} offer 
interesting possibilities for the study of electroweak interaction and for searches 
beyond the Standard Model (\textsc{bsm}) up to the TeV scale. 
The cross sections for various electroweak and \textsc{bsm} 
reactions together with their irreducible (Standard Model) background processes are shown in figure \ref{Fig:gammapartonxsec}.

Several processes have been studied in detail to illustrate the discovery potential of photo-production at \textsc{lhc}. In contrast to photon-photon reactions, the photo-production processes involve topologies with hard jets in the final state. 
The effect of jet algorithms and the efficiency of event selection was taken into account using a fast simulation of a typical multipurpose \textsc{lhc} detector response \cite{severine}. 

The associated $WH$ photo-production cross section reaches 23~$\textrm{fb}$ for a 115~GeV Higgs boson and 17.5~$\textrm{fb}$ for a 170~GeV Higgs boson. It represents more than $2\%$ of the total inclusive $WH$ production at the \textsc{lhc}.

 When the decay branching ratio of the Higgs boson into a $W$ pair becomes dominant, the same sign lepton signature coming from leptonic decays of two out of the three produced $W$ seems promising. It has a signal to noise ratio of about one, which is unique at LHC. An integrated luminosity of a few tens of inverse femptobarns could directly reveal the $HWW$ gauge coupling.

 The $Hbb$ coupling, very challenging to assess in parton-parton processes, could be probed for a light Higgs boson if sufficient integrated luminosity without pile-up can be collected. 

Photo-production of single top is dominated by t-channel amplitudes when the top quark is produced in association with a $W$ boson. In contrast to proton-proton deep inelastic scattering where the ratio of $Wt$ associated production cross section to the sum of all top production cross sections is only about $5\%$, it is about 10 times higher in photo-production as illustrated in Table \ref{VisTop}. This provides a unique opportunity to study this reaction at the start phase of \textsc{lhc}.
 
Cross sections after the application of acceptance cuts are shown in table \ref{xsec_visible}. For the signal, a value of 0.1 was chosen for $k_{tu\gamma}$ while $k_{tc\gamma}$ was set at zero. The resolved $\gamma p \rightarrow Wjq$ is negligible. 

\begin{wraptable}{l}{0.4\columnwidth}
\begin{small}
\begin{center}
\vspace{-0.5cm}
\begin{tabular}{l||c| c}
\hline Cross section [fb] & $\ell bjj$ &  $\ell \ell b$\\\hline
$\sigma \hspace{0.5cm} Wt$                & 440   & 103.7 \\
$\sigma_{acc}$                            & 34.1  & 8.69 \\\hline
\multicolumn{3}{c}{Irreducible processes}\\\hline
$\sigma_{acc}$ $t\overline{t}$            & 46.37 & 2.80\\
$\hspace{0.7cm}Wjjj$                      & 15.61 & -\\
$\hspace{0.7cm}Wb\overline{b}q'$          & 1.01  & -\\
$\hspace{0.7cm}W^+W^-q'$                  &  -    & 0.18  \\\hline
$\sigma_{acc}$ total                      & 62.99 & 2.99\\\hline
\end{tabular}
\end{center}
\caption{\footnotesize Cross sections for two $Wt$ induced final states before and after acceptance cuts
 together with the cross sections of irreducible background processes after acceptance cuts.}\label{VisTop}
\end{small}
\end{wraptable}

 While the overall photoproduction of top quark is sensitive to the, yet unmeasured, top quark electrical charge, the $Wt$ associated photoproduction amplitudes are all proportional to the \textsc{ckm} matrix element $|V_{tb}|$. Preliminary studies show that the di-leptonic channel could be competitive with the inclusive channel as detailed in \cite{severine-top2008}.
Single top photoproduction could reveal phenomena beyond the Standard Model, and in particular Flavour Changing Neutral Currents (\textsc{fcnc})\cite{jerome}.
The effective Lagrangian for this anomalous coupling can be written as \cite{eff_lag_anotop} : 
$$ L = iee_t\bar{t}\frac{\sigma_{\mu\nu}q^{\nu}}{\Lambda}k_{tu\gamma}uA^{\mu} + iee_t\bar{t}\frac{\sigma_{\mu\nu}q^{\nu}}{\Lambda}k_{tc\gamma}cA^{\mu} + h.c., $$
where $\sigma^{\mu\nu}$ is defined as $(\gamma^{\mu} \gamma^{\nu} - \gamma^{\nu} \gamma^{\mu})/2$, $q^{\nu}$ being the photon 4-vector and $\Lambda$ an arbitrary scale, conventionally taken as the top mass. The best upper limit on $k_{tu\gamma}$ is around 0.14, depending on the top mass \cite{zeus_st} while the anomalous coupling $k_{tc\gamma}$ has not been probed yet.

 The final state is composed of a $b$-jet and a $W$ boson. The studied topology is therefore $\ell b$. 
Main irreducible background processes come from $\gamma p$ interactions producing a $W$ boson and a jet mistagged as a $b$-jet.
 
\begin{wraptable}{r}{0.3\columnwidth}
\begin{center}
\begin{small}
\vspace{-.4cm}
\begin{tabular}{l||c}
\hline
Cross section [fb]          & $\ell b$ \\
\hline
$\sigma\hspace{0.4cm}$ $t$  & 769.0               \\
$\sigma_{acc}$              & 144.0              \\
\hline
\multicolumn{2}{c}{Irreducible backgrounds}      \\
\hline
$\sigma_{acc}$ $Wj$           &   56.2             \\
$\hspace{0.7cm}$$Wc$          &   82.8             \\
\hline
$\sigma_{acc}$ total          &  139.0             \\
\hline
\end{tabular}
\caption{\footnotesize Cross sections for one anomalous top induced final state ($k_{tu\gamma}$ = 0.1,
 $k_{tc\gamma}$ = 0) before and after requiring 1 lepton and 1 b-jet within $|\eta|<2.5$ together with the cross sections
 of irreducible background processes after these acceptance cuts.}
\label{xsec_visible}
\end{small}
\end{center}
\end{wraptable}

 The present selection together with the assumption that no other background contribution will interfere, 
would lead to the expectation of a five sigma discovery just below $1~\textrm{fb}^{-1}$ of integrated luminosity. 
The extracted limits on the anomalous couplings k$_{tu\gamma}$ and k$_{tc\gamma}$ are reported in table \ref{cl.anotop}. 
\begin{wraptable}{r}{0.45\columnwidth}
\begin{center}
\vspace{-.8cm}
\begin{small}
\begin{tabular}{c||c|c}
\hline
Coupling & \multicolumn{2}{c}{Limits} \\ 
         &L = 1~fb$^{-1}$ & L = 10~fb$^{-1}$       \\
\hline
k$_{tu\gamma}$  & 0.043 & 0.024                 \\
k$_{tc\gamma}$  & 0.074 & 0.042                 \\
\hline
\end{tabular}
\caption{\footnotesize Expected limits for anomalous couplings at 95$\%$ CL for two values of integrated luminosities.}
\label{cl.anotop}
\end{small}
\end{center}
\end{wraptable}

\section{Conclusion}
A survey of high energy $\gamma\gamma$ and  $\gamma p$ interactions at \textsc{lhc} has been presented. 
The large cross sections show the potential of the study of massive electroweakly interacting particles in a complementary way to the usual, parton-parton processes, at the very start of \textsc{lhc}.

Photon-photon interactions will allow precise luminosity monitoring while 
large integrated luminosity studies open complementary ways to search for \textsc{susy} particles.

At high luminosity, the efficient selection of photon induced 
processes is conditioned by the capacity of experiments to tag forward protons by means of dedicated forward detectors.
In this context, \textsc{r\&d} programs such as \textsc{fp420}\cite{FP420} and other developments 
related to the instrumentation of forward regions of \textsc{cms} and \textsc{atlas} experiments, are particularly important.


\begin{footnotesize}




\begin{thebibliography}{99}
\bibitem{url} Slides: \verb$http://indico.cern.ch/getFile.py/access?contribId=46&sessionId=16&resId=0&$\\ 
\verb$materialId=slides&confId=24657$

\bibitem{EPA}V.M.~Budnev, I.F.~Ginzburg, G.V.~Meledin and V.G.~Serbo, Phys.\ Rept.\  {\bf 15}, 181 (1974).
\bibitem{lumi} X. Rouby, to appear in the proceedings of XLIIIrd Rencontres de Moriond on QCD and hard interactions, arXiv:0805.4406 [hep-ex]
\bibitem{OPAL.limits} G.~Abbiendi {\it et al.}  (OPAL Collaboration), Phys.\ Rev.\  D {\bf 70} (2004) 032005.
\bibitem{Nicolas}N. Schul and K. Piotrkowski, to appear in the proceedings of Workshop on High Energy Photon Collisions at the LHC, Geneva, Switzerland, arXiv:0806.1097 [hep-ph]
\bibitem{severine} S. Ovyn, to appear in the proceedings of Workshop on High Energy Photon Collisions at the LHC, Geneva, Switzerland, arXiv:0806.1157 [hep-ph]
\bibitem{severine-top2008} S. Ovyn and J. de Favereau de Jeneret to appear in the proceedings of Top 2008, International Workshop on Top Physics, arXiv:0806.4841 [hep-ph]
\bibitem{jerome} J. de Favereau de Jeneret and S. Ovyn, to appear in the proceedings of Workshop on High Energy Photon Collisions at the LHC, Geneva, Switzerland, arXiv:0806.4886 [hep-ph]
\bibitem{eff_lag_anotop} T.~Han and J.L.~Hewett, Phys. Rev. D\textbf{60}, (1999) 074015
\bibitem{zeus_st} S.~Chekanov et al., Phys. Lett. B\textbf{559}, (2003) 153-170.
\bibitem{FP420} M.G. Albrow et al.(FP420 R\&D Collaboration),arXiv:0806.0302 [hep-ph]
\end{thebibliography}
%

\end{footnotesize}


\end{document}